\documentclass{article}
\usepackage{amsmath}
\usepackage{amsfonts}
\usepackage{amssymb}
\usepackage[british]{babel}
\usepackage[sans]{dsfont}
\usepackage[superscript]{cite}
\usepackage{epsfig}
\usepackage{amsthm}
\theoremstyle{plain}
\newtheorem{theorem}{Theorem}
\newcommand{\Tr}{\operatorname{Tr}}
\newcommand{\rmd}{\mathrm{d}}
\newcommand{\rmi}{\mathrm{i}}
\newcommand{\rme}{\mathrm{e}}
\newcommand{\openone}{\mathds1}
\newcommand{\norm}[1]{\left\Vert#1\right\Vert}
\newcommand{\abs}[1]{\left\vert#1\right\vert}
\newcommand{\Real}{\mathbb{R}}
\newcommand{\Complex}{\mathbb{C}}

\newcommand{\Pbb}{\mathbb{P}}

\newcommand{\Hscr}{\mathcal{H}}

\newcommand{\Lscr}{\mathcal{L}}

\newcommand{\Sscr}{\mathcal{S}}

 \newcommand{\E}{\operatorname{\mathbb{E}}}
\newcommand{\Cov}{\operatorname{Cov}}
\newcommand{\Var}{\operatorname{Var}}

\newcommand{\IM}{\operatorname{Im}}

\begin{document}

\title{QUANTUM CONTINUAL MEASUREMENTS:\\
THE SPECTRUM OF THE OUTPUT}

\author{A. BARCHIELLI and M. GREGORATTI
\\
\\
Department of Mathematics, Politecnico di Milano,\\
Piazza Leonardo da Vinci, I-20133 Milano, Italy.\\
Also: Istituto Nazionale di Fisica Nucleare, Sezione di Milano. \\
Alberto.Barchielli@polimi.it\qquad
Matteo.Gregoratti@polimi.it\\
www.mate.polimi/QP}

\maketitle

\begin{abstract}
When a quantum system is monitored in continuous time, the result of the measurement is a
stochastic process. When the output process is stationary, at least in the long run, the
spectrum of the process can be introduced and its properties studied. A typical continual
measurement for quantum optical systems is the so called homodyne detection. In this paper we
show how the Heisenberg uncertainty relations give rise to characteristic bounds on the
possible homodyne spectra and we discuss how this is related to the typical quantum phenomenon
of squeezing.
\end{abstract}

\section{Quantum continual measurements}\label{S1}

A big achievement in the 70's-80's was to show that, inside the axiomatic formulation of
quantum mechanics, based on \emph{positive operator valued measures} and
\emph{instruments},\cite{Kra80,Dav76} a consistent formulation of the theory of measurements
continuous in time (\emph{quantum continual measurements}) was
possible.\cite{Dav76,Hol01,BarLP82,BarL85,Bar86PR,Bel88,BarB91} The main applications of
quantum continual measurements are in the photon detection theory in quantum optics
(\emph{direct, heterodyne, homodyne
detection}).\cite{Bar90QO,Car93,WisMil93,WisM93,Wis96,Mil96,GarZ00} A very flexible and
powerful formulation of continual measurement theory was based on stochastic differential
equations, of classical type (commuting noises, It\^o calculus) and of quantum type (non
commuting noises, Hudson-Parthasarathy
equation).\cite{BarL85,Bar86PR,Bel88,BarB91,Bar90QO,Car93,WisMil93,
WisM93,Wis96,Mil96,GarZ00,Bar05}

In this paper we start by giving a short presentation of continual measurement theory based on
quantum SDE's. We consider only the type of observables relevant for the description of
homodyne detection and we make the mathematical simplification of introducing only bounded
operators on the Hilbert space of the quantum system and a finite number of noises. Then, we
introduce the spectrum of the classical stochastic process which represents the output and we
study the general properties of the spectra of such classical processes by proving
characteristic bounds due to the Heisenberg uncertainty principle. Finally, we present the
case of a two-level atom, where the spectral analysis of the output can reveal the phenomenon
of squeezing of the fluorescence light, a phenomenon related to the Heisenberg uncertainty
relations.

\subsection{Hudson Parthasarathy equation}

Let $\Hscr$ be the \emph{system space}, the complex separable Hilbert space associated to the
observed quantum system, which we call system $S$. Quantum stochastic calculus and the
Hudson-Parthasarathy equation\cite{Parthas92} allow to represent the continual measurement
process as an interaction of system $S$ with some quantum fields combined with an observation
in continuous time of these fields. Let us start by introducing such fields. We denote by
$\Gamma$ the Hilbert space associated with $d$ boson fields, that is the symmetric \emph{Fock
space} over the ``one--particle space'' $ L^2(\mathbb{R}_+)\otimes\Complex^d \simeq
L^2({\mathbb R}_+; \Complex^d)$, \ and we denote by \ $e(f)$,  $f \in L^2({\mathbb R}_+;
\Complex^d )$, \ the \emph{coherent vectors}, whose components in the \ $0,1,\ldots,n,\ldots$
\ particle spaces are $ e(f) :=
\exp\left(-\frac{1}{2}\,\|f\|^2\right)\left(1,f,(2!)^{-1/2}f\otimes f,\ldots,(n!)^{-1/2} f^{
\otimes n}, \ldots \right)$.

Let $\{z_k,\ k\geq 1\}$ be the canonical basis in $\Complex^d$ and for any $f\in
L^2(\mathbb{R}_+;\Complex^d)$ let us set $f_k(t):= \langle z_k|f(t)\rangle$. We denote by
$A_k(t)$, $A^\dagger_k(t)$, $\Lambda_{kl}(t)$ the \emph{annihilation, creation and
conservation processes}:
\begin{gather*}
A_k(t)\,e(f)= \int_0^t f_k(s) \,\rmd  s \, e(f)\,, \\
\langle e(g)| A_k^\dagger(t)e(f)\rangle = \int_0^t \overline{ g_k(s)}\, \rmd  s \, \langle
e(g)| e(f)\rangle,
\\
\langle e(g)| \Lambda_{kl}(t)e(f)\rangle = \int_0^t \overline{ g_k(s)}\,f_l(s)\, \rmd  s \,
\langle e(g)| e(f)\rangle.
\end{gather*}
The annihilation and creation processes satisfy the \emph{canonical commutation rules} (CCR);
formally, $[A_k(t),A_l^\dagger(s)]=t\wedge s$, $[A_k(t),A_l(s)]=0$,
$[A_k^\dagger(t),A_l^\dagger(s)]=0$.

Let $H$, $R_k$, $S_{kl}$, $k,l=1,\ldots,d$, be bounded operators on $\Hscr$ such that $H^* =H$
and $\sum_{j}S^*_{jk}S_{jl}=\sum_{j}S_{kj}S^*_{lj}=\delta_{kl}$. We set also $ K:=-\rmi H -
\frac{1}{2} \sum_k R_k^{* }R_k$. Then, the quantum stochastic differential equation
\cite{Parthas92}
\begin{multline}\label{HPequ}
\rmd U(t) = \biggl\{ \sum_k R_k \,\rmd A_k^\dagger(t) + \sum_{kl} \left(S_{kl}-
\delta_{kl}\right) \rmd \Lambda_{kl}(t)
\\ {}- \sum_{kl} R_k^{* } S_{kl }\,\rmd A_l(t) +
K\,\rmd t \biggr\}  U(t),
\end{multline}
with the initial condition $U(0) =\openone$, has a unique solution, which is a strongly
continuous family of unitary operators on $\Hscr\otimes \Gamma$, representing the system-field
dynamics in the interaction picture with respect to the free field evolution.

\subsection{The reduced dynamics of the system}

The states of a quantum system are represented by statistical operators, positive trace-class
operators with trace one; let us denote by $\Sscr(\Hscr)$ the set of statistical operators on
$\Hscr$. As initial state of the composed system ``system $S$ plus fields'' we take
$\rho\otimes \varrho_\Gamma(f)$, where $\rho\in \Sscr(\Hscr)$ is generic and
$\varrho_\Gamma(f)$ is a coherent state, $\varrho_\Gamma(f):= |e(f)\rangle \langle e(f)|$. One
of the main properties of the Hudson-Parthasarathy equation is that, with such an initial
state, the reduced dynamics of system $S$ obeys a quantum master
equation.\cite{Parthas92,Bar05} Indeed, we get
\begin{equation}
\frac{\rmd \ }{\rmd t}\, \eta_t=\Lscr(t)[\eta_t], \qquad \eta_t:=\Tr_\Gamma \left\{ U(t)\bigl(
\rho \otimes \varrho_\Gamma(f)\bigr)U(t)^*\right\},
\end{equation}
where the Liouville operator $\Lscr(t)$ turns out to be given by
\begin{multline}
\Lscr(t)[\rho]= \left(K-\sum_{kl}R_k^* S_{kl}f_l(t)\right)\rho + \rho
\left(K^*-\sum_{kj}\overline{f_j(t)}S_{kj}^{\;*}R_k\right)
\\ {} +
\sum_k \left(R_k-\sum_l S_{kl}f_l(t)\right)\rho\left(R_k^*-
S_{kl}^{\;*}\overline{f_l(t)}\right) - \norm{f(t)}^2\rho.
\end{multline}
A particularly important case is $S_{kl}=\delta_{kl}$, when $\Lscr(t)$ reduces to
\begin{multline}
\Lscr(t)[\rho]= -\rmi\left[ H-\rmi\sum_k f_k(t)R_k^*+ \rmi\sum_k\overline{f_k(t)}R_k,\,\rho
\right]
\\ {} +
\sum_k \left(R_k\rho R_k^*-\frac 1 2 R_k^*R_k\rho -\frac 1 2\rho R_k^*R_k\right).
\end{multline}

It is useful to introduce also the evolution operator from $s$ to $t$ by
\begin{equation}
\frac{\rmd \ }{\rmd t}\,\Upsilon(t,s)=\Lscr(t)\circ \Upsilon(t,s),\qquad
\Upsilon(s,s)=\openone.
\end{equation}
With this notation we have $ \eta_t=\Upsilon(t,0)[\rho] $.

\subsection{The field observables}

The key point of the theory of continual measurements is to consider field observables
represented by time dependent, commuting families of selfadjoint operators in the Heisenberg
picture.\cite{Bar05} Being commuting at different times, these observables represent outputs
produced at different times which can be obtained in the same experiment. Here we present a
very special case of family of observables, a field quadrature. Let us start by introducing
the operators
\begin{equation}\label{quadrature}
Q(t;\vartheta,\nu)= \int_0^t \rme^{-\rmi \left(\nu s +\vartheta\right)} \rmd A_1^\dagger(s) +
\int_0^t \rme^{\rmi \left(\nu s +\vartheta\right)} \rmd A_1(s), \qquad t\geq 0;
\end{equation}
$\vartheta\in (-\pi,\pi]$ and $ \nu >0$ are fixed. The operators $Q(t;\vartheta,\nu)$ are
selfadjoint (they are essentially selfadjoint on the linear span of the exponential vectors).
By using CCR's, one can check that they commute: $[Q(t;\vartheta,\nu),Q(s;\vartheta,\nu)]=0$
(better: the unitary groups generated by $Q(t;\vartheta,\nu)$ and $Q(s;\vartheta,\nu)$
commute). The operators \eqref{quadrature} have to be interpreted as linear combinations of
the formal increments $\rmd A_1^\dagger(s) $, $\rmd A_1(s) $ which represent field operators
evolving with the free-field dynamics; therefore, they have to be intended as operators in the
interaction picture. The important point is that these operators commute for different times
also in the Heisenberg picture, because
\begin{equation}\label{Qout}
Q^{\mathrm{out}}(t;\vartheta,\nu):= U(t)^*Q(t;\vartheta,\nu)U(t)=
U(T)^*Q(t;\vartheta,\nu)U(T), \quad\forall T\geq t;
\end{equation}
this is due to the factorization properties of the Fock space and to the properties of the
solution of the Hudson-Parthasarathy equation. These ``output'' quadratures are our
observables. They regard those bosons in ``field 1'' which eventually have interacted with $S$
between time $0$ and time $t$. Commuting selfadjoint operators can be jointly diagonalized and
the usual postulates of quantum mechanics give the probabilities for the joint measurement of
the observables represented by the selfadjoint operators $Q^{\mathrm{out}}(t;\vartheta,\nu)$,
$t\geq 0$. Let us stress that operators of type \eqref{quadrature} with different angles and
frequencies represent incompatible observables, because they do not commute but satisfy
\[
[Q(t;\theta,\nu),Q(s;\phi,\mu)]=\frac{4\rmi\sin\left(\frac{t\wedge s}2
\left(\nu-\mu\right)\right)\sin \left(\theta-\phi+\frac{t\wedge s}2
\left(\nu-\mu\right)\right)}{\nu-\mu}\,  .
\]

When ``field 1'' represents the electromagnetic field, a physical realization of a measurement
of the observables \eqref{Qout} is implemented by what is called an heterodyne/homodyne
scheme. The light emitted by the system in the ``channel 1'' interferes with an intense
coherent monochromatic laser beam of frequency $\nu$. The mathematical description of the
apparatus is given in Section 3.5 of Ref.\ \citen{Bar05}.

\subsection{Characteristic operator, probabilities and moments}

The commuting selfadjoint operators \eqref{quadrature} have a joint pvm (projection valued
measure) which gives the distribution of probability for the measurement. Anyway, at least the
finite-dimensional distributions of the output can be obtained via an explicit and easier
object, the \emph{characteristic operator} $\widehat \Phi_t(k;\vartheta,\nu)$, a kind of
Fourier transform of this pvm. For any real test function $k \in L^\infty(\Real_+)$ and any
time $t>0$ we define the unitary Weyl operator
\begin{multline}\label{WQh}
\widehat \Phi_t(k;\vartheta,\nu) = \exp \biggl\{ \rmi  \int_0^t k(s)\, \rmd Q(s;\vartheta,\nu)
\biggr\}
\\ {}=
\exp \biggl\{ \rmi  \int_0^t k(s) \rme^{-\rmi \left(\nu s +\vartheta\right)} \rmd
A_1^\dagger(s) + \rmi \int_0^t k(s) \rme^{\rmi \left(\nu s +\vartheta\right)} \rmd A_1(s)
\biggr\}.
\end{multline}
Then, there exists a measurable space $(\Omega, \mathcal{F})$, a pvm
$\xi_\vartheta^\nu$ (acting on $\Gamma$) with value space $(\Omega, \mathcal{F})$, a family of
real valued measurable functions $\left\{ X(t;\cdot)\,, \; t\geq 0\right\}$ on $\Omega$, such
that $ X(0;\omega)=0$, and, for any choice of $n$, $0=t_0<t_1<\cdots < t_n \leq t$,
$\kappa_j\in \Real$,
\begin{multline}\label{Phiincrements}
\widehat \Phi_t(k;\vartheta,\nu) = \exp\biggl\{ \rmi \sum_{j=1}^n  \kappa_j \big[
Q(t_j;\vartheta,\nu) - Q(t_{j-1};\vartheta,\nu)\big]\biggr\}
\\
{}= \int_\Omega \exp\biggl\{ \rmi \sum_{j=1}^n  \kappa_j \big[ X(t_j;\omega) -
X(t_{j-1};\omega)\big]\biggr\} \xi_\vartheta^\nu(\rmd \omega)\,,
\end{multline}
where $k(s)= \sum_{j=1}^n 1_{(t_{j-1},t_j)}(s) \,\kappa_j$. Let us stress that the pvm depends
on the observables and, so, on the parameters $\vartheta$ and $\nu$, while the choices of the
trajectory space (the measurable space $(\Omega, \mathcal{F})$) and of the process $X(t)$ are
independent of $\vartheta$ and $\nu$.

Then, we introduce the characteristic functional
\begin{multline}\label{characteristicf}
\Phi_t(k;\vartheta,\nu) = \Tr \left\{\exp \biggl\{ \rmi  \int_0^t k(s)\, \rmd
Q^{\mathrm{out}}(s;\vartheta,\nu) \biggr\} \rho\otimes \varrho_\Gamma(f) \right\}
\\ {}
= \Tr \left\{\widehat\Phi_t(k;\vartheta,\nu)U(t) \left(\rho\otimes
\eta(f)\right)U(t)^*\right\}.
\end{multline}
All the finite-dimensional probabilities of the increments of the process $X(t)$ are
determined by
\begin{multline}\label{fdcf}
\Phi_t(k;\vartheta,\nu) = \int_{\Real^{n}} \biggl(\prod_{j=1}^n  \rme^{\rmi\kappa_j \cdot
x_j}\biggr)
\\
{}\times \Pbb_{\rho}^{\vartheta,\nu}\big[\Delta X(t_0,t_1)\in \rmd x_1 ,\ldots, \Delta
X(t_{n-1},t_n)\in \rmd x_n\big]\,,
\end{multline}
where we have  introduced the test function $k(s)= \sum_{j=1}^n 1_{(t_{j-1},t_j)}(s)
\,\kappa_j$, with $0=t_0<t_1<\cdots < t_n \leq t$, $\kappa_j\in \Real$.

The fact that the theory gives in a simple direct way the distribution for the increments of the process
$X(t)$,
rather than its finite-dimensional distributions, is related also to the
interpretation: the output $X(t)$ actually is obtained by a continuous observation of
the generalized process $I(t)= \rmd X(t)/\rmd t$ followed by post-measurement processing.

Starting from the characteristic functional it is possible to obtain the moments of the output
process $I(t)$ and to express them by means of quantities concerning only system
$S$.\cite{Bar05} Let us denote by $\E_\rho^{\vartheta,\nu}$ the expectation with respect to
$\Pbb_\rho^{\vartheta,\nu}$; for the first two moments we obtain the expressions
\begin{subequations}\label{moments}
\begin{equation}
\E_\rho^{\vartheta,\nu}[I(t)]=\Tr\left\{\left(Z(t)+Z(t)^*\right)\eta_t\right\},
\end{equation}
\begin{multline}
\E_\rho^{\vartheta,\nu}[I(t)I(s)] = \delta(t-s)
\\ {}+\Tr\left\{\left(Z(t_2)+Z(t_2)^*\right)\Upsilon(t_2,t_1)\left[Z(t_1)\eta_{t_1}+\eta_{t_1}
Z(t_1)^*\right] \right\},
\end{multline}
where $t_1=t\wedge s$, $t_2=t\vee s$ and
\begin{equation}
Z(t) := \rme^{\rmi(\nu t +\vartheta)}\left(R_1+\sum_k S_{1k}f_k(t)\right).
\end{equation}
\end{subequations}

\section{The spectrum of the output}

\subsection{The spectrum of a stationary process}
In the classical theory of stochastic processes, the spectrum is related to the Fourier
transform of the autocorrelation function. Let $Y$ be a stationary real stochastic process
with finite moments; then, the mean is independent of time $ \E[Y(t)]=\E[Y(0)]=:m_Y$, $\forall
t\in \Real$, and the second moment is invariant under time translations
\begin{equation}\label{2_st}
\E[Y(t)Y(s)]=\E[Y(t-s)Y(0)]=:R_Y(t-s), \qquad \forall t,s\in \Real\,.
\end{equation}
The function $R_Y(\tau)$, $\tau\in \Real$, is called the \emph{autocorrelation function} of
the process. Obviously, we have $ \Cov\left[Y(t),Y(s)\right]=R_Y(t-s)-m_Y^{\,2}$.

The \emph{spectrum} of the stationary stochastic process $Y$ is the Fourier transform of its
autocorrelation function:
\begin{equation}
S_Y(\mu):=\int_{-\infty}^{+\infty}\rme^{\rmi \mu \tau}R_Y(\tau)\, \rmd \tau\,.
\end{equation}
This formula has to be intended in the sense of distributions. For instance, if
$\Cov\left[Y(\tau),Y(0)\right]\in L^1(\Real)$, we can write
\begin{equation}
S_Y(\mu):=2\pi m_Y^{\,2} \delta(\mu)+\int_{-\infty}^{+\infty}\rme^{\rmi \mu
\tau}\Cov\left[Y(\tau),Y(0)\right] \rmd \tau\,.
\end{equation}
By the properties of the covariance, the function $\Cov\left[Y(\tau),Y(0)\right]$ is positive
definite and, by the properties of positive definite functions, this implies
\[
\int_{-\infty}^{+\infty}\rme^{\rmi \mu \tau}\Cov\left[Y(\tau),Y(0)\right] \rmd \tau\geq 0;
\]
then, also $S_Y(\mu)\geq 0$.

By using the stationarity and some tricks on multiple integrals, one can check that an
alternative expression of the spectrum is
\begin{equation}\label{eq:sp2}
S_Y(\mu)= \lim_{T\to + \infty}\frac 1 T \E\left[\abs{\int_0^T  \rme^{\rmi \mu t}Y(t)\,\rmd t
}^2\right].
\end{equation}
The advantage now is that positivity appears explicitly and only positive times are involved.
Expression \eqref{eq:sp2} can be generalized also to processes which are stationary only in
some asymptotic sense and to singular processes as our $I(t)$.

\subsection{The spectrum of the output in a finite time horizon}
Let us consider our output $I(t)= \rmd X(t)/\rmd t$ under the physical probability
$\Pbb_\rho^{\vartheta,\nu}$. We call ``spectrum up to time $T$'' of $I(t)$ the quantity
\begin{equation}\label{eq:spI(t)}
S_T(\mu;\vartheta,\nu)= \frac 1 T \E_\rho^{\vartheta,\nu}\left[\abs{\int_0^T  \rme^{\rmi \mu
t}\,\rmd X(t) }^2\right].
\end{equation}
When the limit $T\to +\infty$ exists, we can speak of \emph{spectrum of the output}, but this
existence depends on the specific properties of the concrete model.

By writing the second moment defining the spectrum as the square of the mean plus the
variance, the spectrum splits in an elastic or coherent part and in an inelastic or incoherent
one:
\begin{equation}
S_T(\mu;\vartheta,\nu)=
S_T^{\mathrm{el}}(\mu;\vartheta,\nu)+S_T^{\mathrm{inel}}(\mu;\vartheta,\nu),
\end{equation}
\begin{equation}
S_T^{\mathrm{el}}(\mu;\vartheta,\nu)= \frac 1 T \abs{\E_\rho^{\vartheta,\nu}\left[\int_0^T
\rme^{\rmi \mu t}\,\rmd X(t)\right] }^2,
\end{equation}
\begin{multline}
S_T^{\mathrm{inel}}(\mu;\vartheta,\nu)= \frac 1 T \Var_\rho^{\vartheta,\nu}\left[\int_0^T
\cos\mu t\,\rmd X(t) \right] \\ {}+ \frac 1 T\Var_\rho^{\vartheta,\nu}\left[\int_0^T \sin\mu
t\,\rmd X(t) \right] ,
\end{multline}
Let us note that
\begin{equation}
S_T^{\mathrm{el}}(\mu;\vartheta,\nu)=S_T^{\mathrm{el}}(-\mu;\vartheta,\nu), \qquad
S_T^{\mathrm{inel}}(\mu;\vartheta,\nu)=S_T^{\mathrm{inel}}(-\mu;\vartheta,\nu).
\end{equation}

By using the expressions \eqref{moments} for the first two moments we get the spectrum in a
form which involves only system operators:
\begin{subequations}\label{spectrumS}
\begin{equation}
S_T^{\mathrm{el}}(\mu;\vartheta,\nu)= \frac 1 T \abs{\int_0^T \rme^{\rmi \mu t}\Tr \left\{
\left(Z(t)+Z(t)^*\right) \eta_t\right\}\rmd t}^2,
\end{equation}
\begin{multline}
S_T^{\mathrm{inel}}(\mu;\vartheta,\nu) = 1  +\frac 2 T \int_0^T\rmd t  \int_0^t \rmd s
\,\cos\mu(t-s)
\\ {}\times \Tr\left\{\left(\tilde Z(t)+\tilde Z(t)^*\right)\Upsilon(t,s)\left[\tilde
Z(s)\eta_s+\eta_s \tilde Z(s)^*\right] \right\},
\end{multline}
\begin{equation}
\tilde Z(t) = Z(t) - \Tr\left\{Z(t) \eta_t\right\}.
\end{equation}
\end{subequations}

\subsection{Properties of the spectrum and the Heisenberg uncertainty relations}

Equations \eqref{spectrumS} give the spectrum in terms of the reduced description of system
$S$ (the fields are traced out); this is useful for concrete computations. But the general
properties of the spectrum are more easily obtained by working with the fields; so, here we
trace out first system $S$. Let us define the reduced field state
\begin{equation}
\Pi_T(f):=\Tr_\Hscr\left\{U(T)\bigl( \rho \otimes \varrho_\Gamma(f)\bigr)U(T)^*\right\}
\end{equation}
and the field operators
\begin{subequations}
\begin{gather}
Q_T(\mu;\vartheta,\nu)= \frac 1 {\sqrt{T}}\int_0^T\rme^{\rmi \mu t}\,\rmd Q(t;\vartheta,\nu),
\\
\tilde Q_T(\mu;\vartheta,\nu)= Q_T(\mu;\vartheta,\nu)-\Tr\left\{\Pi_T(f)Q_T(\mu;\vartheta,\nu)
\right\}.
\end{gather}
\end{subequations}
Let us stress that $Q_T(\mu;\vartheta,\nu)$ commutes with its adjoint and that
$Q_T(\mu;\vartheta,\nu)^*=Q_T(-\mu;\vartheta,\nu)$. By using Eqs.\ \eqref{characteristicf} and
\eqref{fdcf} and taking first the trace over $\Hscr$, we get
\begin{subequations}\label{eq:sp2I(t)}
\begin{gather}
S_T(\mu;\vartheta,\nu)= \Tr\left\{\Pi_T(f)
Q_T(\mu;\vartheta,\nu)^*Q_T(\mu;\vartheta,\nu)\right\}\geq 0,
\\
S_T^{\mathrm{el}}(\mu;\vartheta,\nu)=  \abs{ \Tr\left\{\Pi_T(f)
Q_T(\mu;\vartheta,\nu)\right\}}^2\geq 0,
\\
S_T^{\mathrm{inel}}(\mu;\vartheta,\nu)=\Tr\left\{\Pi_T(f) \tilde
Q_T(\mu;\vartheta,\nu)^*\tilde Q_T(\mu;\vartheta,\nu)\right\}\geq 0.
\end{gather}
\end{subequations}

To elaborate the previous expressions it is useful to introduce annihilation and creation
operators for bosonic modes, which are only approximately orthogonal for finite $T$:
\begin{subequations}
\begin{equation}
a_T(\omega):= \frac 1 {\sqrt{T}} \int_0^T \rme^{\rmi \omega t} \rmd A_1(t)= \frac {\rme^{\frac
\rmi  2\, \omega T}} {\sqrt{T}}\int_{-\frac T 2}^{\frac T 2} \rme^{\rmi \omega t} \rmd
A_1(t+T/2),
\end{equation}
\begin{gather}
[a_T(\omega), a_T(\omega^\prime)]= [a_T^\dagger(\omega), a_T^\dagger(\omega^\prime)]=0,
\\
[a_T(\omega), a_T^\dagger(\omega^\prime)]= \begin{cases} 1 & \text{for } \omega^\prime=\omega,
\\
\frac{\rme^{\rmi (\omega -\omega^\prime)T}-1}{ \rmi  (\omega -\omega^\prime)T} & \text{for }
\omega^\prime\neq \omega.
\end{cases}
\end{gather}
\end{subequations}
Then, we have easily
\begin{equation}\label{twomodequadrature}
Q_T(\mu;\vartheta,\nu)=\rme^{\rmi \vartheta} a_T(\nu+\mu)+\rme^{-\rmi \vartheta}
a_T^\dagger(\nu-\mu),
\end{equation}
\begin{subequations}\label{thetadependence}
\begin{multline}
S_T(\mu;\vartheta,\nu)=1 + \Tr\Bigl\{\Pi_T(f) \Bigl( a_T^\dagger(\nu+\mu)a_T(\nu+\mu)
+a_T^\dagger(\nu-\mu)a_T(\nu-\mu) \\ {}+\rme^{-2\rmi \vartheta}
a_T^\dagger(\nu+\mu)a_T^\dagger(\nu-\mu) +\rme^{2\rmi \vartheta}
a_T(\nu-\mu)a_T(\nu+\mu)\Bigr)\Bigr\},
\end{multline}
\begin{equation}
S_T^{\mathrm{el}}(\mu;\vartheta,\nu)=\abs{\rme^{\rmi \vartheta}\Tr\left\{\Pi_T(f)
a_T(\nu+\mu)\right\} + \rme^{-\rmi \vartheta}\Tr\left\{\Pi_T(f)
a_T^\dagger(\nu-\mu)\right\}}^2.
\end{equation}
\end{subequations}

\begin{theorem}
Independently of the system state $\rho$, of the field state $\varrho_\Gamma(f)$ and of the
Hudson-Parthasarathy evolution $U$, for every $\vartheta$ and $\nu$ we have the two bounds
\begin{gather}
\frac 1 2 \left(S_T^{\mathrm{inel}}(\mu;\vartheta,\nu) + S_T^{\mathrm{inel}}(\mu;\vartheta\pm
{\textstyle \frac\pi 2},\nu)\right) \geq 1,
\\ \label{Heisenberg}
S_T^{\mathrm{inel}}(\mu;\vartheta,\nu)S_T^{\mathrm{inel}}(\mu;\vartheta\pm {\textstyle \frac
\pi 2},\nu) \geq 1.
\end{gather}
\end{theorem}

\begin{proof} The first bound comes easily from
\[
S_T^{\mathrm{inel}}(\mu;\vartheta,\nu) = S_T(\mu;\vartheta,\nu)-
S_T^{\mathrm{el}}(\mu;\vartheta,\nu)
\]
and Eqs.\ \eqref{thetadependence}.

To prove the second bound, let us introduce the operator
\begin{subequations}
\begin{equation}
B_T(\omega):= a_T(\nu-\omega) - \Tr \left\{ \Pi_T(f)a_T(\nu-\omega)\right\},
\end{equation}
which satisfy the CCR
\begin{equation}
[B_T(\omega), B_T^\dagger(\omega)]= 1, \qquad [B_T(\omega), B_T(\omega)]=
[B_T^\dagger(\omega), B_T^\dagger(\omega)]=0.
\end{equation}
\end{subequations}
Then, we can write
\begin{multline*}
S_T^{\mathrm{inel}}(\mu;\vartheta,\nu)= \Tr\Bigl\{\left( \rme^{-\rmi \vartheta}
B_T^\dagger(-\mu) + \rme^{\rmi \vartheta} B_T(\mu)\right) \Pi_T(f)\\ {} \times\left(
\rme^{-\rmi\vartheta} B_T^\dagger(\mu) +\rme^{\rmi \vartheta} B_T(-\mu)\right) \Bigr\}.
\end{multline*}

The usual tricks to derive the Heisenberg-Scr\"odinger-Robertson uncertainty relations can be
generalized also to non-selfadjoint operators.\cite{Hol01,CavesS1} For any choice of the state
$\varrho$ and of the operators $X_1$, $X_2$ (with finite second moments with respect to
$\varrho$) the $2\times 2$ matrix with elements $ \Tr\left\{X_i\varrho X_j^*\right\}$ is
positive definite and, in particular, its determinant is not negative. Then, we have
\begin{multline*}
\Tr\left\{X_1\varrho X_1^*\right\}\Tr\left\{X_2\varrho X_2^*\right\}\geq
\abs{\Tr\left\{X_1\varrho X_2^*\right\}}^2 \\ {}\geq \abs{\IM\Tr\left\{X_1\varrho
X_2^*\right\}}^2 = \frac 1 4 \abs{\Tr\left\{\varrho\left( X_2^*X_1-X_1^*X_2\right)\right\}}^2.
\end{multline*}
By taking \qquad $\varrho =\Pi_T(f)$, \qquad $X_1=\rme^{-\rmi \vartheta} B_T^\dagger(\mu)
+\rme^{\rmi \vartheta} B_T(-\mu)$, \qquad \\ $X_2=\mp \rmi \left(\rme^{-\rmi \vartheta}
B_T^\dagger(\mu) -\rme^{\rmi \vartheta} B_T(-\mu)\right)$, \quad we get
\begin{multline*}
S_T^{\mathrm{inel}}(\mu;\vartheta,\nu)S_T^{\mathrm{inel}}(\mu;\vartheta\pm {\textstyle \frac
\pi 2},\nu)
\\ {}\geq \abs{1 + \Tr\left\{\Pi_T(f) \left( B_T^\dagger(\mu)B_T(\mu)
-B_T^\dagger(-\mu)B_T(-\mu)\right)\right\}}^2.
\end{multline*}
But we can change $\mu$ in $-\mu$ and we have also
\begin{multline*}
S_T^{\mathrm{inel}}(\mu;\vartheta,\nu)S_T^{\mathrm{inel}}(\mu;\vartheta\pm {\textstyle
\frac\pi 2},\nu) =S_T^{\mathrm{inel}}(-\mu;\vartheta,\nu)S_T^{\mathrm{inel}}(-\mu;\vartheta\pm
{\textstyle \frac\pi 2},\nu)
\\ {}\geq \abs{1 + \Tr\left\{\Pi_T(f) \left( B_T^\dagger(-\mu)B_T(-\mu)
-B_T^\dagger(\mu)B_T(\mu)\right)\right\}}^2.
\end{multline*}
The two inequalities together give
\begin{multline}
S_T^{\mathrm{inel}}(\mu;\vartheta,\nu)S_T^{\mathrm{inel}}(\mu;\vartheta\pm {\textstyle
\frac\pi 2},\nu)
\\ {}\geq \left(1 + \abs{\Tr\left\{\Pi_T(f) \left( B_T^\dagger(\mu)B_T(\mu)
-B_T^\dagger(-\mu)B_T(-\mu)\right)\right\}}\right)^2\geq 1,
\end{multline}
which is what we wanted.
\end{proof}

Ref.\ \citen{CavesS1} introduces a class of operators for the electromagnetic field, called
\emph{two-mode quadrature-phase amplitudes}, which have the structure
\eqref{twomodequadrature} of our operators $Q_T(\mu;\vartheta,\nu)$. Anyway only two modes are
involved, as if we fixed $\mu $ and $\nu$. Let us denote here those operators by
$Q_\mathrm{pa}$. The paper explicitly constructs a class of quasi-free (or Gaussian) field
states $\varrho_\mathrm{sq}$ for which $\Tr \left\{\varrho_\mathrm{sq}
Q_\mathrm{pa}^*Q_\mathrm{pa}\right\}- \abs{\Tr \left\{\varrho_\mathrm{sq}
Q_\mathrm{pa}\right\}}^2<1$. Such states are called \emph{two-mode squeezed states}. More
generally, one speaks of \emph{squeezed light} if, at least in a region of the $\mu$ line, for
some $\vartheta$ one has $S_T^{\mathrm{inel}}(\mu;\vartheta,\nu)<1$. If this happens, the
Heisenberg-type relation \eqref{Heisenberg} says that necessarily
$S_T^{\mathrm{inel}}(\mu;\vartheta+\frac \pi 2,\nu)>1$ in such a way that the product is
bigger than one.

\section{Squeezing of the fluorescence light of a two-level atom}
Let us take as system $S$ a two-level atom, which means $\Hscr=\Complex^2$, $H=\frac
{\omega_0} 2\, \sigma_z$; $\omega_0>0$ is the \emph{resonance frequency} of the atom. We
denote by $\sigma_-$ and $\sigma_+$ the lowering and rising operators and by
$\sigma_x=\sigma_-+\sigma_+$, $\sigma_y=\rmi(\sigma_--\sigma_+)$, $\sigma_z=\sigma_+\sigma_- -
\sigma_-\sigma_+$ the Pauli matrices; we set also $ \sigma_\vartheta =
\rme^{\rmi\vartheta}\,\sigma_- + \rme^{-\rmi\vartheta}\,\sigma_+ $. We stimulate the atom with
a coherent monochromatic laser and consider homodyne detection of the fluorescence light. The
quantum fields $\Gamma$ model the whole environment. The electromagnetic field is split in two
fields, according to the direction of propagation: one field for the photons in the forward
direction ($k=2$), that of the stimulating laser and of the lost light, one field for the
photons collected to the detector ($k=1$). Assume that the interaction with the atom is
dominated by absorption/emission and that the direct scattering is negligible:
\[
S_{kl}=\delta_{kl}\,, \qquad R_1=\sqrt{\gamma p}\, \sigma_-\,, \qquad R_2=\sqrt{\gamma
(1-p)}\, \sigma_-\,.
\]
The coefficient $\gamma>0$ is the natural \emph{line-width} of the atom, $p$ is the fraction
of fluorescence light which reaches the detector and $1-p$ is the fraction of lost light
($0<p<1$).\cite{Car93,GarZ00,Bar05,BarGL08} We introduce also the interaction with a thermal
bath,
\[
R_3=\sqrt{\gamma \overline{n}}\, \sigma_-\,, \qquad R_4=\sqrt{\gamma \overline{n}}\,
\sigma_+\,, \qquad \overline{n}\geq 0,
\]
and a term responsible of \emph{dephasing} (or decoherence),
\[
R_5=\sqrt{\gamma k_d}\, \sigma_z\,, \qquad k_d\geq 0.
\]

To represent a coherent monochromatic laser of frequency $\omega>0$, we take $f_k(t)=
\delta_{k2}\, \frac{\rmi \Omega} {2\sqrt{\gamma (1-p)}}\, \rme^{-\rmi \omega t}1_{[0,T]}(t)$;
$T$ is a time larger than any other time in the theory and the limit $T\to +\infty$ is taken
in all the physical quantities. The quantity $\Omega\geq 0$ is called \emph{Rabi frequency}
and $\Delta\omega=\omega_0-\omega$ is called \emph{detuning}. The squeezing in the
fluorescence light is revealed by homodyne detection, which needs to maintain phase coherence
between the laser stimulating the atom and the laser in the detection apparatus which
determines the observables $Q(t;\vartheta,\nu)$; this in particular means that necessarily we
must take $\nu=\omega$.

The limit $T\to +\infty$ can be taken in Eqs.\ \eqref{spectrumS} and it is independent of the
atomic initial state.\cite{BarGL08} The result is
\begin{equation}
S^{\mathrm{el}}(\mu;\vartheta):=\lim_{T\to + \infty} S_T^{\mathrm{el}}(\mu;\vartheta,\omega)=2
\pi \gamma p \abs{\Tr\left\{\sigma_\vartheta \rho_\textrm{eq}\right\}}^2  \delta(\mu),
\end{equation}
\begin{equation}
S^{\mathrm{inel}}(\mu;\vartheta):= \lim_{T\to + \infty}
S_T^{\mathrm{inel}}(\mu;\vartheta,\omega)=1+2\gamma
p\,\left(\frac{A}{A^2+\mu^2}\,\vec{t}\right)\cdot\vec{s},
\end{equation}
where
\begin{gather*}
\vec{t}=\Tr\Big[\big(\rme^{\rmi\vartheta}\sigma_-\,\rho_\textrm{eq} +
\rho_\textrm{eq}\,\rme^{-\rmi\vartheta}\sigma_+ -
\Tr[\sigma_{\vartheta}\,\rho_\textrm{eq}]\,\rho_\textrm{eq}\big)\,\vec\sigma\Big],\qquad
\vec{s}=\begin{pmatrix}\cos\vartheta\\ \sin\vartheta\\0\end{pmatrix},
\\
\rho_\textrm{eq}=\frac{1}{2}\left(1+\vec{x}_\textrm{eq}\cdot\vec\sigma\right), \qquad
\vec{x}_\textrm{eq} = -\gamma A^{-1}\,
\begin{pmatrix}0\\0\\1\end{pmatrix},
\\
A=\begin{pmatrix}\gamma\left(\frac{1}{2}+\overline{n}+2k_\rmd \right)& \Delta \omega&0\\
-\Delta \omega &\gamma\left(\frac{1}{2}+\overline{n}+2k_\rmd \right)&\Omega\\
0&-\Omega&\gamma\left(1+2\overline{n} \right)\end{pmatrix}.
\end{gather*}

\begin{figure}[b]
\begin{center}
\parbox{5.2cm}{
\psfig{file=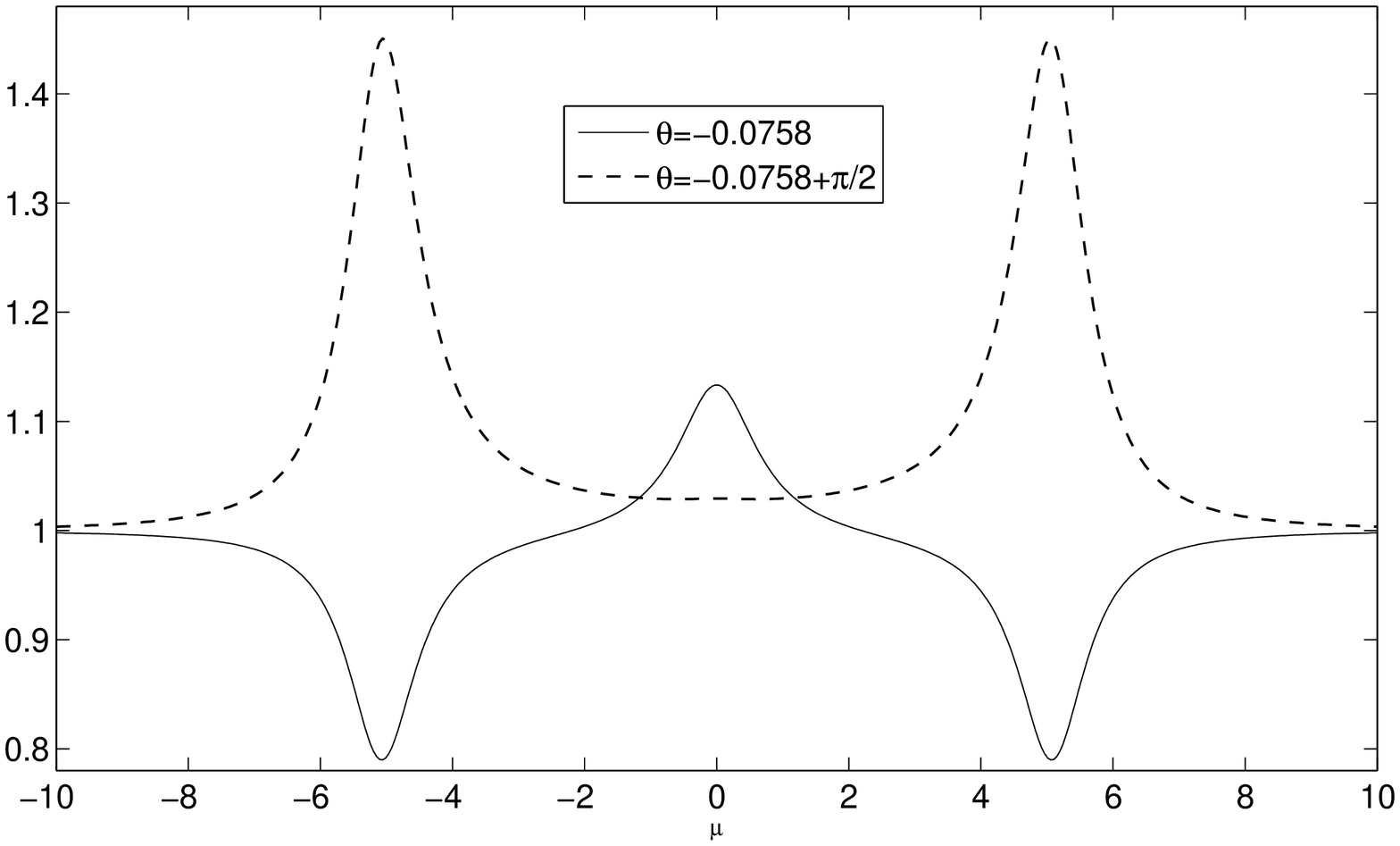,height=4cm,width=5.1cm} \caption{$S^{\mathrm{inel}}(\mu;\vartheta)$
with  $\Delta\omega=3.5$,  $\Omega=3.7021$.} \label{fig1}} \quad
\begin{minipage}{5.2cm}
\psfig{file=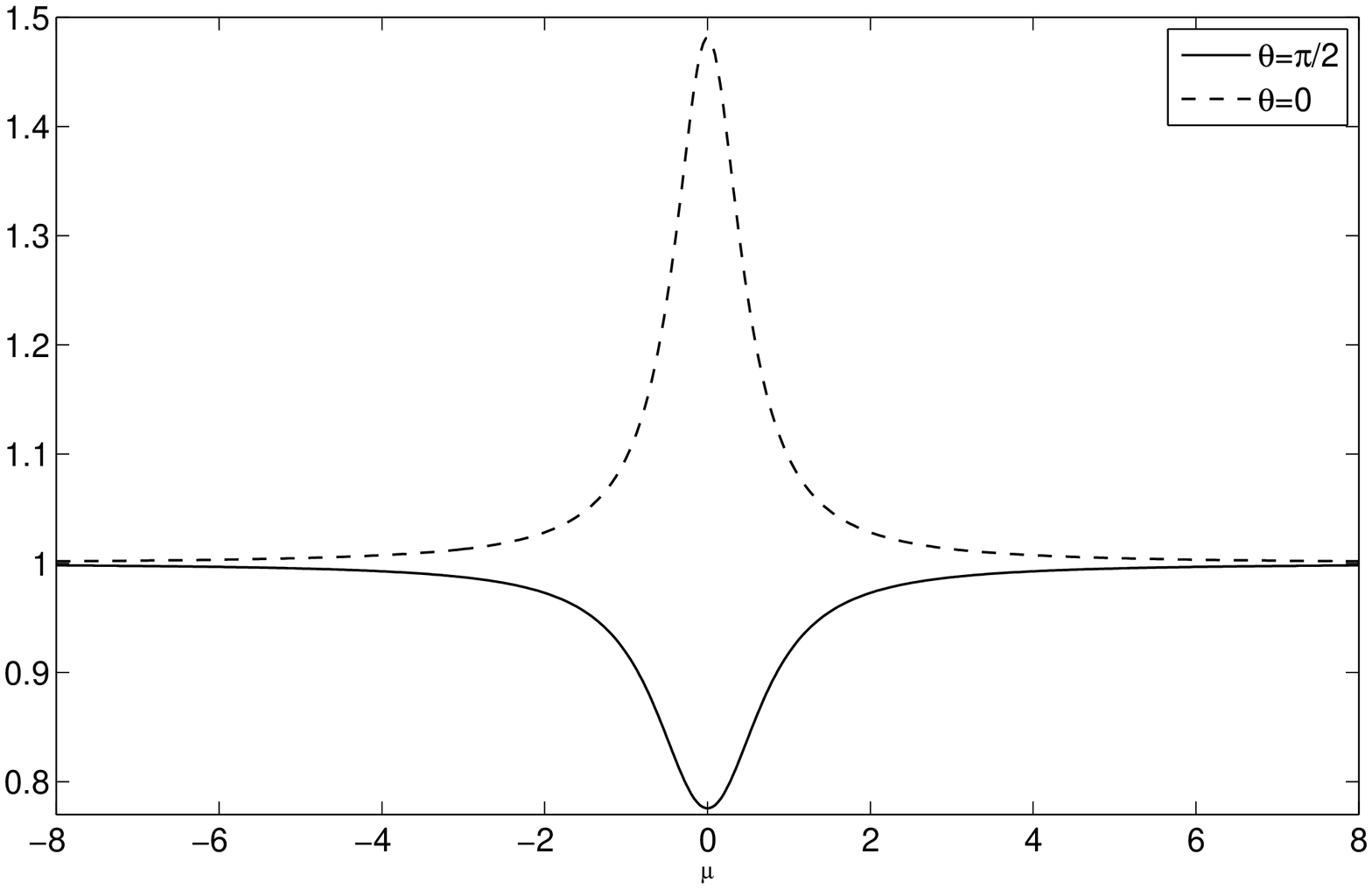,height=4cm,width=5.1cm} \caption{$S^{\mathrm{inel}}(\mu;\vartheta)$
with  $\Delta\omega=0$,  $\Omega=0.2976$.} \label{fig2}
\end{minipage}
\end{center}
\end{figure}
Examples of inelastic spectra are plotted for $\gamma=1$, $\overline{n}=k_\rmd=0$, $p=4/5$,
and two different values of $\Delta\omega$. The Rabi frequency $\Omega$ and $\theta$ are
chosen in both cases to get good visible minima of $S^{\mathrm{inel}}$ below 1. Thus in this
case the analysis of the homodyne spectrum reveals the squeezing of the detected light. Also
complementary spectra are shown to verify Theorem 2.1. One could also compare the homodyne
spectrum with and without $\overline n$ and $k_\rmd$, thus verifying that the squeezing is
very sensitive to any small perturbation.

\end{document}